**Formation of a Nematic Fluid at High Fields in $Sr_3Ru_2O_7$**

R.A. Borzi [1,2,*], S.A. Grigera [1,3], J. Farrell [1], R.S. Perry [1], S.J.S. Lister [1], S.L. Lee [1], D.A. Tennant [4], Y. Maeno [5] & A.P. Mackenzie [1,*]

[1] *Scottish Universities Physics Alliance, School of Physics and Astronomy, University of St.Andrews, North Haugh, St.Andrews, Fife KY16 9SS, UK.*

[2] *Instituto de Investigaciones Fisicoquímicas Teóricas y Aplicadas (UNLP-CONICET), c.c. 16, Suc. 4, 1900 La Plata, Argentina, and Departamento de Física, IFLP, UNLP, c.c. 67, 1900 La Plata, Argentina.*

[3] *Instituto de Física de Líquidos y Sistemas Biológicos (UNLP-CONICET-CIC), c.c.565, 1900 La Plata, Argentina.*

[4] *Hahn-Meitner-Institut (HMI), Glienicker Str. 100, D-14109 Berlin, Germany*

[5] *Department of Physics and International Innovation Center, Kyoto University, Kyoto 606-8501, Japan.*

*\* To whom correspondence should be addressed to r.chufo@gmail.com or apm9@st-and.ac.uk*

**Abstract**

In principle, a complex assembly of strongly interacting electrons can self-organise into a wide variety of collective states, but relatively few such states have been identified in practice. We report that, in the close vicinity of a metamagnetic quantum critical point, high purity $Sr_3Ru_2O_7$ possesses a large magnetoresistive anisotropy, consistent with the existence of an electronic nematic fluid. We discuss a striking phenomenological similarity between our observations and those made in high purity two-dimensional electron fluids in GaAs devices.

In the standard materials that form the basis of most of today's electronic technology, the Hamiltonian for the outer electrons is dominated by the attraction to the ions of the crystalline lattice. In 'strongly correlated' materials this is no longer true. The Coulomb interaction between electrons is large, so might be expected to add a large term to the Hamiltonian that is not necessarily strongly related to the periodic potential. However, in most correlated electron systems studied to date, the many-electron collective states still retain strong links with the lattice, and the range of 'correlated electron matter' identified so far is considerably less diverse than should, in principle, be possible. In itinerant systems, it almost always consists of electron liquids such as Fermi liquids or superfluids that respect the lattice symmetry; the identification of superconductors in which the condensate breaks some lattice symmetries has been one of the triumphs of the field *(1)*.

In recent years, there have been proposals that even more exotic electronic liquids might be observable. In an analogy with the nematic state of liquid crystals, which is characterised by orientational but not positional order, it might be possible to form nematic liquids in electronic systems with strong correlations *(2)*. In the broadest sense, a correlated electron nematic is characterised by a lowering of rotational symmetry in its itinerant properties that is not simply a consequence of a symmetry lowering of the lattice.

We report electrical transport phenomena which show that the correlated electron system possesses this key property of a nematic fluid. In previous work, we have argued that a novel quantum phase forms in the vicinity of a metamagnetic quantum critical point in the correlated electron oxide $Sr_3Ru_2O_7$ *(3-5)*. Here, we show that this state is accompanied by pronounced magnetoresistive anisotropies which have two-fold symmetry and can be aligned using modest in-plane magnetic fields. Even in the presence of these two-fold anisotropies, neutron single crystal diffraction resolves no change from the initial square symmetry of the lattice. The overall phenomenology of our observations bears a striking resemblance to that observed in GaAs devices near the high-field limit *(6-8)*, suggesting that the nematic phenomena previously thought to be special to close proximity to a Fractional Quantum Hall State may be more general.

The single crystals used in the present work were grown in an image furnace using techniques described fully in ref. *(9)*. All transport data shown are measurements of the in-plane magnetoresistivity, denoted $\rho$ *(10)*. Crystal purity is crucial in $Sr_3Ru_2O_7$. For a residual resistivity $\rho_o \approx 3$ μΩcm (corresponding to a mean free path of approximately 300 Å or less) the phase diagram contains a quantum critical point (QCP) which can be accessed by the application of a magnetic field of approximately 7.8 T parallel to the crystalline *c* axis, and whose effects have now been studied using a variety of experimental probes *(11-16)*. As the purity is increased, first-order phase transitions appear as the QCP is approached, and measurements of at least five thermodynamic and transport properties contain features whose loci enclose a well-defined region of the phase diagram in the vicinity of the QCP. Previously, we have argued that these observations are indicative of the formation of a new quantum phase *(3-5)*. For fields applied parallel to *c*, $\rho$ has a pronounced anomaly when this phase is entered (Fig. 1A). The two steep 'sidewalls' coincide with first order phase transitions that can be observed using a.c. susceptibility or magnetization. The angle, $\theta$, of the applied magnetic field to the *ab* plane of the crystal is a known tuning parameter in $Sr_3Ru_2O_7$ *(13)*. Previous work has shown that the large resistive anomaly of Fig. 1A disappears rapidly with tilt angle *(17)*, leaving behind much weaker signals in $\rho$ which trace the origin of the two first-order phase transitions as a bifurcation from a single first-order transition at $\theta \approx 60°$ *(5)*, marked by the white arrow in Fig. 2A. At first sight, this seems to contradict the identification of the bounded region between the two first order lines as a single distinctive phase – there was no evidence that the sudden drop in resistivity with angle at $\theta \approx 80°$ coincided with any phase boundary. The apparent contradiction can be resolved by postulating the existence of domains of some kind. In such a picture, the behaviour shown in Figs. 1A, 2A would be due to these domains producing the extra scattering. The fact that the anomalous scattering disappears so rapidly with angle would then be most straightforwardly interpreted as being due to the in-plane component of the tilted magnetic field ($H_{\text{in-plane}}$) destroying the domains.

**Figure 1** *The two diagonal components $\rho_{aa}$ and $\rho_{bb}$ of the in-plane magnetoresistivity tensor of a high purity single crystal of $Sr_3Ru_2O_7$. (A) For an applied field parallel to the crystalline c axis (with an alignment accuracy of better than 2°), $\rho_{aa}$ (black) and $\rho_{bb}$ (red) are almost identical. (B) With the crystal tilted such that the field is 13° from c, giving an in-plane component along a, a pronounced anisotropy is seen, with the 'easy' direction for current flow being along b, perpendicular to the in-plane field component (17). If the in-plane field component is aligned along b instead, the easy direction switches to being for current flow along a.*

However, instead of simply removing the anomalous peak, $H_{\text{in-plane}}$ exposed an intrinsic asymmetry of the underlying phase, defining 'easy' and 'hard' directions for magnetotransport. These easy and hard directions are shown in Fig. 1B. In previous experiments *(3-5)*, we had worked with the current $I \parallel b \perp H_{\text{in-plane}}$, so that the standard metallic transverse magnetoresistance $\rho_{bb}$ could be studied across the whole phase diagram. In that configuration (red traces in Fig. 1), the anomalous scattering disappears rapidly as $H_{\text{in-plane}}$ increases. However, the behaviour of $\rho_{aa}$ (measured with $I \parallel a \parallel H_{\text{in-plane}}$) is completely different. As shown by the black traces in Fig. 1, the scattering rate remains high even for an angle at which the anomalous scattering is absent for $I \perp H_{\text{in-plane}}$.

Pronounced in-plane resistive anisotropy can have a number of origins. There is known to be a strong magneto-structural coupling in $Sr_3Ru_2O_7$, so one possibility is a symmetry-lowering structural phase transition giving the resistive anisotropy due to a corresponding anisotropy in the hopping integrals. Another is the formation of field-alignable magnetic domains, such as those examined in the context of itinerant metamagnetism in recent theoretical work *(18, 19)*.

To investigate these possibilities we first carried out measurements of $\rho(\mathbf{H})$ and magnetic susceptibility $\chi(\mathbf{H})$ at temperatures between 20 mK and 4 K, on 20 samples from three different batches with a wide variety of shapes, 6 of which were cut from the same piece of crystal (SOM 3). Shapes vary from square plates with sides $A \approx B \gg C$ to rectangular plates with side $A \gg B \gg C$ to long cylinders with $A \approx B \ll C$. In each class of sample, dimension $C$ is aligned with the crystalline $c$ axis, but we deliberately tested combinations of $A$ and $B$ which were both aligned and misaligned with the crystalline $a$ and $b$ axes (SOM 1). An important result of all these experiments is that, apart from a small region of hysteresis (maximum width approximately 80 mT), all the first-order phase boundaries observed in $Sr_3Ru_2O_7$ were invariant under changes to the sample shape. This firmly rules out demagnetization effects as playing a major role in determining the physics, and, hence, directly contradicts predictions concerning magnetic domains *(19)*. To check for a large spontaneous lattice parameter anisotropy, we performed elastic neutron scattering measurements for $H // c$ (SOM 2). Within our experimental resolution of $4 \times 10^{-5}$ Å, we see no evidence of any difference in lattice parameters $a$ and $b$ in the anomalous region.

The experiments described above therefore rule out two of the more standard explanations (magnetic domains and structural change) for the anisotropy shown in Fig. 1B. In Fig. 2 we show the complete field-angle phase diagrams for fields between 4.2 T and 9 T rotating the entire 90º from parallel to $a$ to parallel to $c$ ($\theta = 90º$). The temperature is 100 mK. Fig. 2A shows the magnetoresistivity for the easy direction, $\rho_{bb}$, and Fig. 2B that for the hard direction, $\rho_{aa}$.

*Figure 2. 3-D plots of the magnetoresistivity components $\rho_{bb}$ (A) and $\rho_{aa}$ (B) of a single crystal of $Sr_3Ru_2O_7$ as the external magnetic field is rotated from alignment along the crystalline a axis (0º) to along the crystalline c axis (90º), at a constant temperature of*

100 mK. The quantity $H_c(\theta)$ that normalises h is the main metamagnetic transition (i.e. the one that dominates the change in the magnetic moment). It varies smoothly from 5.1 T at 0° to 7.87 T at 90°. The same data plotted without this normalisation are shown in SOM 3.

Fig. 3 shows typical data for the difference in $\rho$ between the hard and easy directions (main figure) and the temperature dependence of $\rho$ in a field of 7.4 T along both directions (upper inset). Along the easy direction standard metallic behaviour is seen, while a non-metallic temperature dependence is seen for the hard direction (in agreement with results reported previously *(3, 4)*). In the presence of such temperature-dependent anisotropy, it is reasonable to plot the difference between $\rho_{aa}$ and $\rho_{bb}$, normalised by their sum, as a phenomenological order parameter. This is done in the lower inset to Fig. 3 at a field applied at $\theta = 72°$.

**Figure 3. Main panel:** *The temperature dependence of the difference between $\rho_{aa}$ and $\rho_{bb}$ for fields applied at $\theta = 72°$ such that the in-plane field component lies along a.* **Upper inset:** *The temperature dependence of $\rho_{aa}$ (black) and $\rho_{bb}$ (red) for $\mu_oH = 7.4$ T applied in the direction specified above. For this field orientation, $\rho_{bb}$ has a clearly metallic temperature dependence, while $\rho_{aa}$ has the mildly non-metallic temperature dependence previously reported for H // c in refs. (3, 4). If the in-plane field component is instead oriented along b, $\rho_{bb}$ and $\rho_{bb}$ switch in both magnitude and temperature dependence.* **Lower inset:** *The temperature dependence of the difference between the two magnetoresistivities shown in the upper inset, normalised by their sum, which is*

*similar to that expected for the order parameter associated with a continuous thermal phase transition.*

Several striking features are evident from our data: (i) A key finding is that strong scattering can be observed in the whole region of the phase diagram enclosed by the two first-order phase boundaries that bifurcate from a single first order transition line at $\theta \approx 60º$ *(5)*. If there is an in-plane field component, this scattering becomes strongly anisotropic. The present observations are highly significant because if one interprets the lower inset of Fig. 3 as evidence for an order parameter, one sees that the phase that it describes is bounded (at low temperatures and constant $\theta$) by well-defined first-order phase transitions which are independent of extrinsic parameters such as sample shape. (ii) Although the easy direction in Fig. 2 is for currents passed along the *b* axis, this is determined by the fact that the in-plane field component was directed along *a* for the data shown. If the direction of the field is rotated by 90º the anisotropy is reversed and *a* becomes the easy direction. Checks show that the rotation is not smooth; the easy direction is either along *a* or *b* but cannot be made to lie, for example, at 45º. (iii) As can be seen in the lower inset to Fig. 3, the presence of a (small) symmetry breaking $H_{\text{in-plane}}$ slightly rounds off the transition in temperature, giving a 'tail' above 800 mK. This effect, which highlights the fact that an in-plane field (that breaks rotational symmetry) is conjugate to the order parameter, becomes more pronounced for lower angles, that is, for higher $H_{\text{in-plane}}$. (iv) Another feature, weaker, broader but still very noticeable, is seen in the hard direction for $\theta < 40º$ and $h \approx 1.2$ [see Fig. 2B]. (v) The anisotropy described in point (iv) is, like the one described in point (i), extremely sensitive to sample purity (see

*(20)* and SOM 4), strongly suggesting a common origin for the two. Its breadth in field and in temperature (not shown) is also consistent with its being close to the *ab* plane, i.e. in the presence of a large in-plane magnetic field. Specific heat data taken cooling down at its central field show a logarithmic divergence of *C/T* down to 1 K *(20)*, giving good evidence that this feature, like that for fields parallel to *c*, is related to incipient quantum criticality.

The combination of susceptibility, neutron scattering and transport data gives strong evidence for the spontaneous formation of a structured, anisotropic state in the correlated electron fluid as a quantum critical point is approached in $Sr_3Ru_2O_7$. Although the anisotropy is seen explicitly in the presence of a weak symmetry-breaking in-plane field component, there is compelling evidence that the symmetry breaking of the correlated electron state is spontaneous, and exists even for *H // c*. The data shown in Fig. 1A can be reconciled with those at other parts of the phase diagram (Fig. 1B, Fig. 2) if the hard axis is randomly oriented along the *a* and *b* directions in different regions of the sample, leading to overall isotropy in spite of strong local anisotropy. Such behaviour is commonplace in symmetry-broken states, e.g. in simple ferromagnets in zero applied magnetic field *(21)*.

Interesting comparisons can be made between the present data and those in other correlated systems. In-plane transport anisotropies have been observed in both cuprates and manganites. In those systems the crystals are always orthorhombic, but in some cases the anisotropy increases while the degree of orthorhombicity decreases, and strong anomalies have been seen in the Hall effect, both of which have been interpreted as evidence for spontaneous charge stripe formation *(22-24)*. A much stronger similarity exists between our observations and those on 2D GaAs devices. For example, it was shown *(6-8)* that if the devices could be prepared with ultra-high mobility such that the Fractional Quantum Hall effect (FQHE) could be observed in the upper two Landau levels, the correlated electron system does not make a simple FQHE-Fermi liquid crossover as the field is reduced and the filling increased. In the $N$=2 to $N$=5 Landau levels, the FQHE is replaced by a strong spontaneous resistive anisotropy aligned with principal in-plane crystal axes, even though the crystal symmetry shows no evidence of orthorhombicity. The anisotropy exists even for fields perpendicular to the plane of the device (presumably because of some symmetry-breaking gradients introduced during device fabrication) but can be rotated by 90º by applying a modest in-plane field. Just as in the present observations on $Sr_3Ru_2O_7$, the GaAs data have strong temperature and purity dependences, and the easy direction lies perpendicular to an in-plane field applied along one of the two relevant crystalline principal axes.

The phenomenological similarity between the GaAs and $Sr_3Ru_2O_7$ results suggests a common origin for the observations. The disorder dependence gives an important clue, as strong sensitivity to elastic scattering is the signature of a state that is anisotropic in **k**-space, as is well known in unconventional superconductivity *(25)*. The challenge is how to reconcile what are, apparently, large differences in the starting physical situations. In order to promote self-organisation of a correlated electron system, one must tune the ratio of a potential energy term often summarized by the parameter *U* (the Coulomb repulsion which tends to localise) to the kinetic energy, often denoted by $W$, which tends to delocalize. If this is done 'chemically', by forming new compounds, the change in the $U/W$ ratio is strongly linked to a change to the electron-lattice coupling, as increasing $U$ and decreasing $W$ also involves increasing the strength of the periodic potential. In the GaAs devices, it is possible to increase the $U/W$ ratio by quenching the kinetic energy by going to very low Landau levels. This leads to a relatively high effective correlation strength without an increase in the effective strength of the periodic potential. In $Sr_3Ru_2O_7$, similar basic physics is taking place but using a different kind of tuning. It is intrinsically a strongly correlated material, so the starting periodic potential is much larger than in GaAs. However, the existence of an underlying metamagnetic quantum critical point makes the quasiparticle mass $m^*$ diverge on the approach to criticality *(14, 26)*. This mass divergence is another route to increasing $U/W$ without increasing the strength of the periodic potential, hence freeing the correlated electron fluid from its rigid link to the underlying lattice.

We have shown in this paper that in highly restricted parts of its phase diagram, in close proximity to metamagnetic quantum critical points, the electron fluid in $Sr_3Ru_2O_7$ develops a strong resistive anisotropy, whose 'hard' and 'easy' axes can be interchanged by the application of modest in-plane magnetic fields. The data are consistent with the formation of a 'nematic' state with broken rotational symmetry. Intriguingly, a correlated electron nematic arising from a Pomeranchuk-like Fermi surface distortion *(4, 27-35)* possesses two of the key features that are present in our data and those from GaAs, namely the **k**-space anisotropy that would give a strong disorder dependence and the possibility of anisotropic transport, intrinsic or through domain formation *(36)*. Whatever the detailed microscopic origin *(37)*, our data suggest that nematic behaviour is a feature of ultra-clean low-dimensional correlated electron systems in which the bandwidth can be reduced independently of changes to the strength of the periodic potential.

higher noise levels than those reported in, for example, ref. (6). In particular, the small resistive step at 7.8 T for $H_{\text{in-plane}} \perp I$ is harder to resolve than in our high-resolution data.

38. We thank L. Balents, E. Fradkin, A.G. Green, C.A. Hooley, K.Y. Kee, Y.B. Kim, S.A. Kivelson and B.D. Simons for a number of insightful discussions, and the UK EPSRC, Royal Society and Leverhulme Trust for financial support.

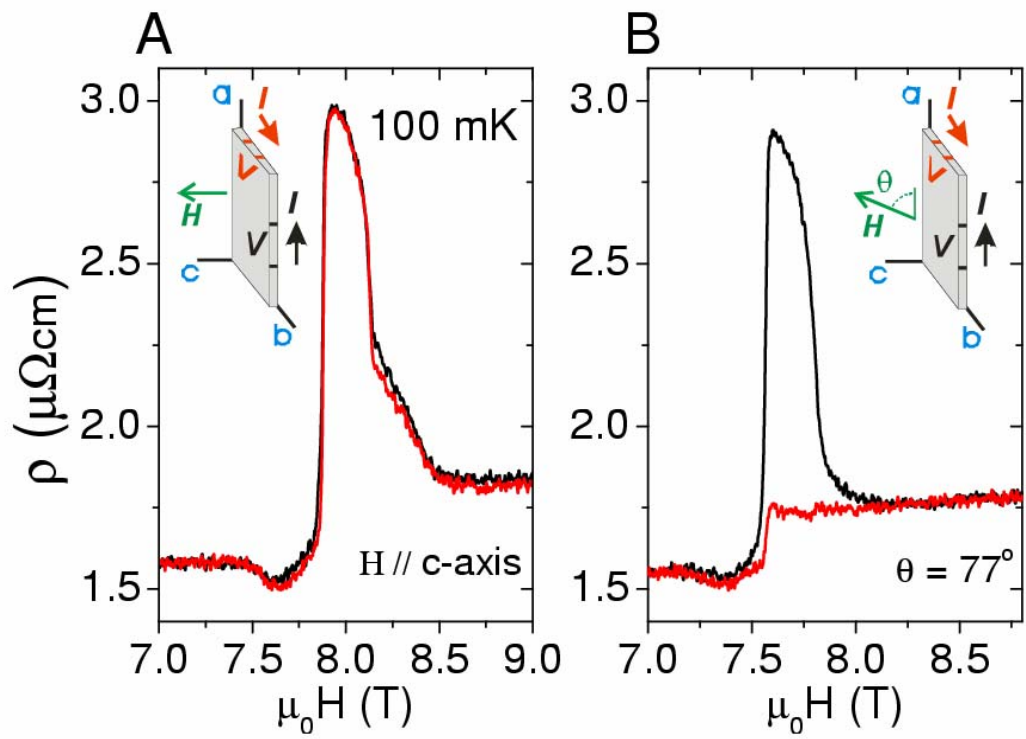

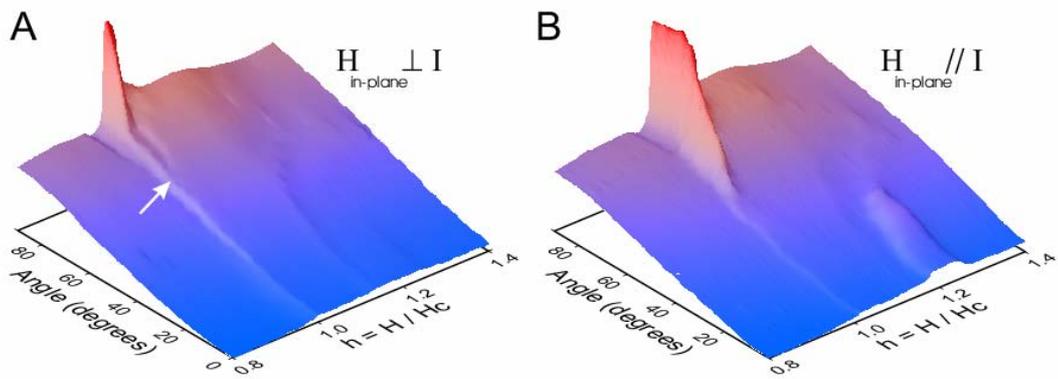

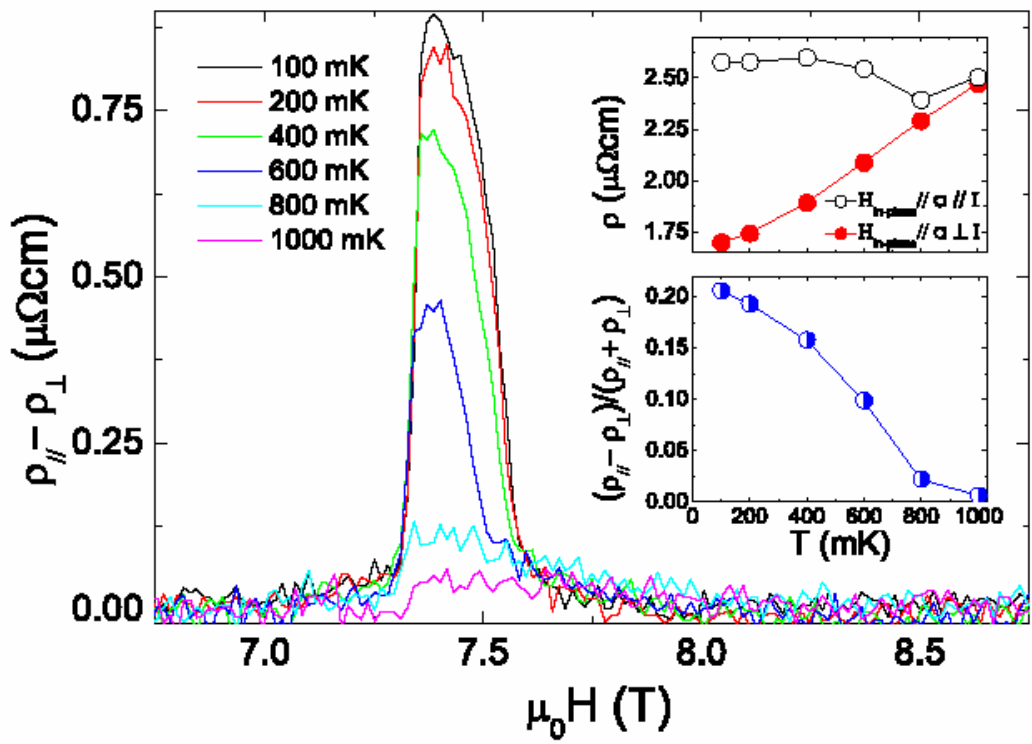

SOM 1

a. Tests for demagnetization effects using magnetic susceptibility

Since 'standard' magnetic domain formation is intimately related to demagnetizing fields, it was important for us to check for possible demagnetization effects on the empirically determined 'phase diagram'. To do this, we first measured magnetic susceptibility ($\chi$) on samples ranging from plates to cylinders. In Fig S1.1 we show the result of two such measurements on samples with approximate dimension along a, b and c-axis, 0.8 x 0.8 x 0.4 (mm$^3$) and 0.8 x 0.8 x 2.5 (mm$^3$) respectively. The magnetic field was approximately aligned along the samples crystallographic c-axis; in this configuration, we estimate the difference in demagnetization factor of the two samples to be approximately a factor of four, but to within experimental error, the main peaks in both $\chi$' and $\chi$'' are separated by identical field ranges. This is in direct contradiction to the clear prediction made in ref. [14] about the formation of Condon-like magnetic domains, where the expectation would be of a factor four difference in the separation of the second and third peaks between the two cases. We also studied the full angular dependence of $\chi$ for the two samples (not shown) and could show that all the main 1$^{st}$ order phase boundaries (the single one for field angle $\theta > 60º$ and the birfurcated ones for $\theta < 60º$) are invariant under this large change in demagnetizing factor.

b. Demagnetisation tests by comparison of resistivity and magnetic susceptibility.

We are able to make even larger changes to the relative demagnetisation factors by comparing resistivity data from thin platelets to susceptibility data from our best approximation to a cylinder. In Fig. S1.2 we show data from 100 mK from panel c of fig. S1.1 compared with the resistivity of a sample with approximate dimensions 2.5 x 0.2 x 0.1 (mm$^3$) along a, b and c-axis, respectively. The demagnetization factors differ by a factor of 5-6, but again, the features that can be identified with first order phase boundaries are separated by the same field range to within experimental error.

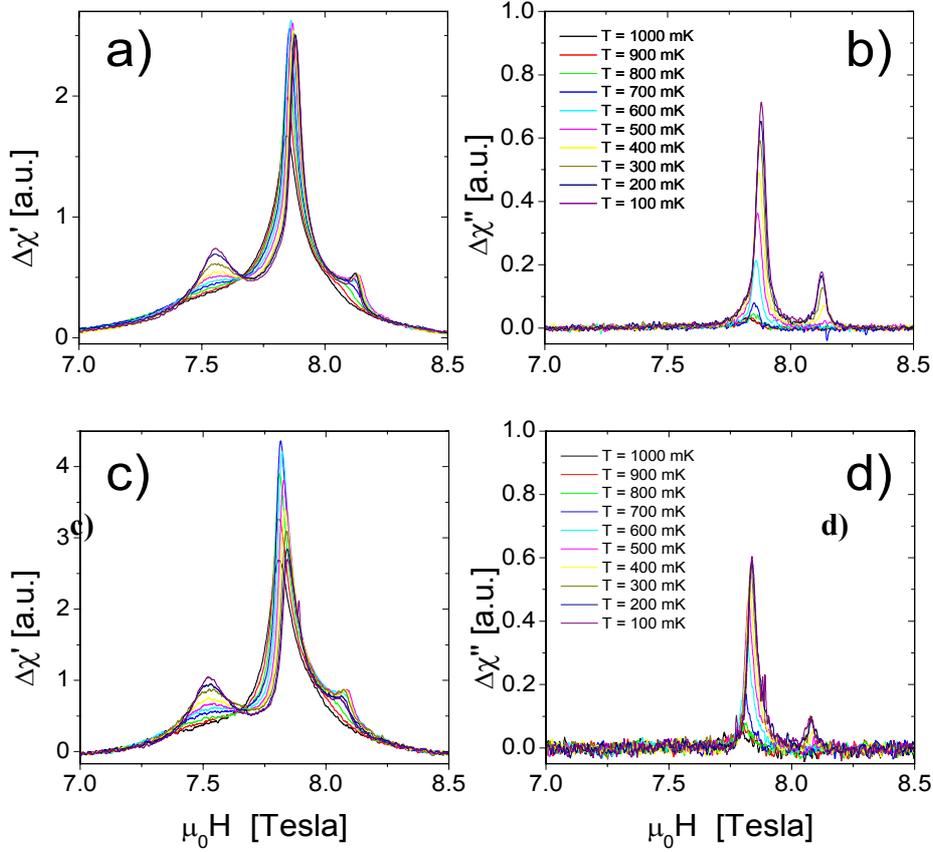

***Figure S1.1*** *Real and imaginary susceptibility χ' and χ'' for two samples, a platelike crystal of dimensions 0.8 x 0.8 x 0.4 (mm$^3$) (panels a and b) and a cylinder of approximate dimensions 0.8 x 0.8 x 2.5 (mm$^3$) (panels c and d). All the metamagnetic peaks are offset by a small amount due to a sample misalignment of a few degrees. However, the main point of the test is clear: within our experimental resolution, the main peaks in susceptibility have the same spacing in magnetic field, even though there is approximately a factor of 4 difference in the demagnetizing factor between the two samples. The data are quoted as Δχ because a smooth background has been subtracted [10].*

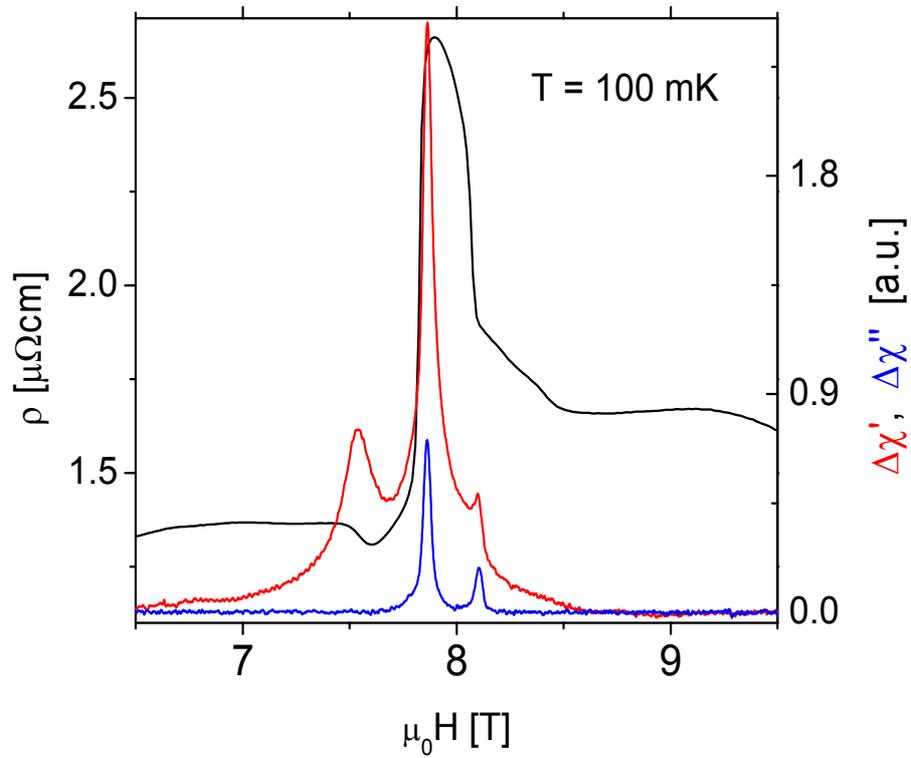

*Figure S1.2* Susceptibility data at 100mK from a cylinder shaped sample with a radius of 0.8 mm and 2.5 mm long compared with resistivity date taken at the same temperature from a thin platelet of approximate dimensions 2.5 x 0.2 x 0.1 mm$^3$ and similar purity.

**SOM 2**

A combination of neutron and capacitive dilatometry measurements show that the lattice parameters in the *a, b*, and *c* directions increase monotonically with an applied magnetic field parallel to the *c*-axis. Consequently the peak shaped anomaly in the resistivity cannot be attributed to a homogeneous change in the magnitude of the hopping integrals.

Given the existence of a strong magneto-structural coupling in $Sr_3Ru_2O_7$, it is important to check whether the anisotropic resistivity discussed in the paper is accompanied by (or due to) the onset of a large structural anisotropy. We have investigated this using elastic neutron scattering at the Hahn-Meitner Institute in Berlin. The required conditions (high applied fields and dilution refrigerator temperatures) are demanding, and do not allow simultaneous observation of diffraction peaks that separately check the lengths of *a* and *b* for a tilted sample. However, it is possible to perform the check by an alternative means.

If the anisotropic resistivity in tilted fields is due to the alignment of domains with a strong structural anisotropy ($a \neq b$), then a reflection giving information about, say, the length of *a* would either broaden or split at the onset of the domain-filled state, because it would probe *two* unequal real space lengths, *a'* from one class of domain and *b'* from the other.

In fig. S2.1 we show data from the (020) reflection of $Sr_3Ru_2O_7$ with the magnetic field applied parallel to *c*. We observe a change of $\Delta a/a \sim 1.2 \times 10^{-4}$ at 8T, which is similar to the relative change in *c* measured using dilatometry [20]. We do *not*, however, resolve any broadening of the measured line-shape for fields in the anomalous region between 7.8 and 8.1 T. We see no sign of such broadening, meaning that *a* = *b* to within 7 parts in $10^6$, with the limit determined by our instrumental resolution.

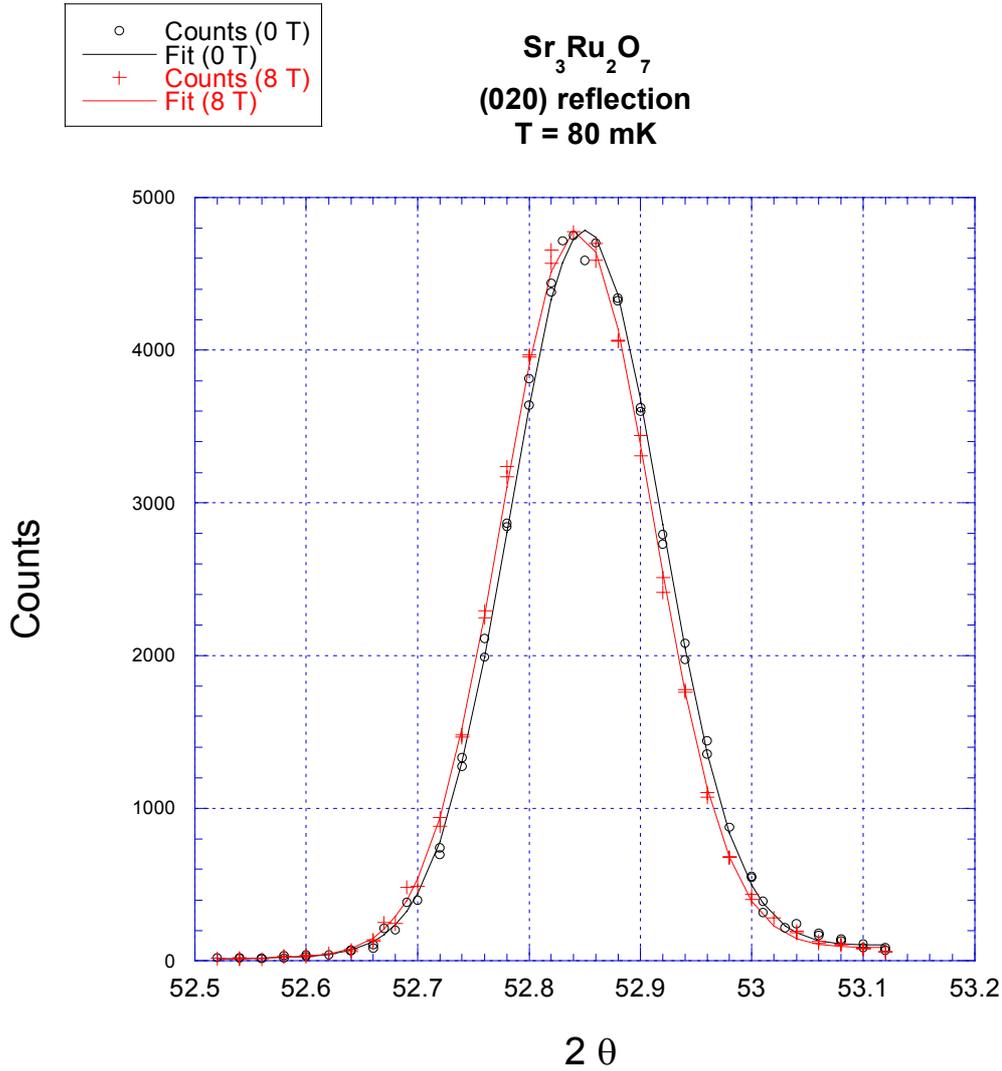

***Figure S2.1*** *Elastic neutron scattering peak from the 020 reflection of a high quality single crystal of $Sr_3Ru_2O_7$ in zero applied magnetic field (black) and a field of 8 tesla applied parallel to c (red). The neutron wavelength was 2.428 Å, so the shift in the peak value corresponds to a change in the in-plane lattice parameter from 5.45586 Å to 5.45653 Å. The relative change of $1.2 \times 10^{-4}$ is very similar to that observed in the c parameter using dilatometry [20]. The width of the peaks remains constant within experimental uncertainty, allowing us to place an upper limit of $4 \times 10^{-5}$ Å on the difference between the a and b lattice parameters in the anomalous region.*

**SOM 3**

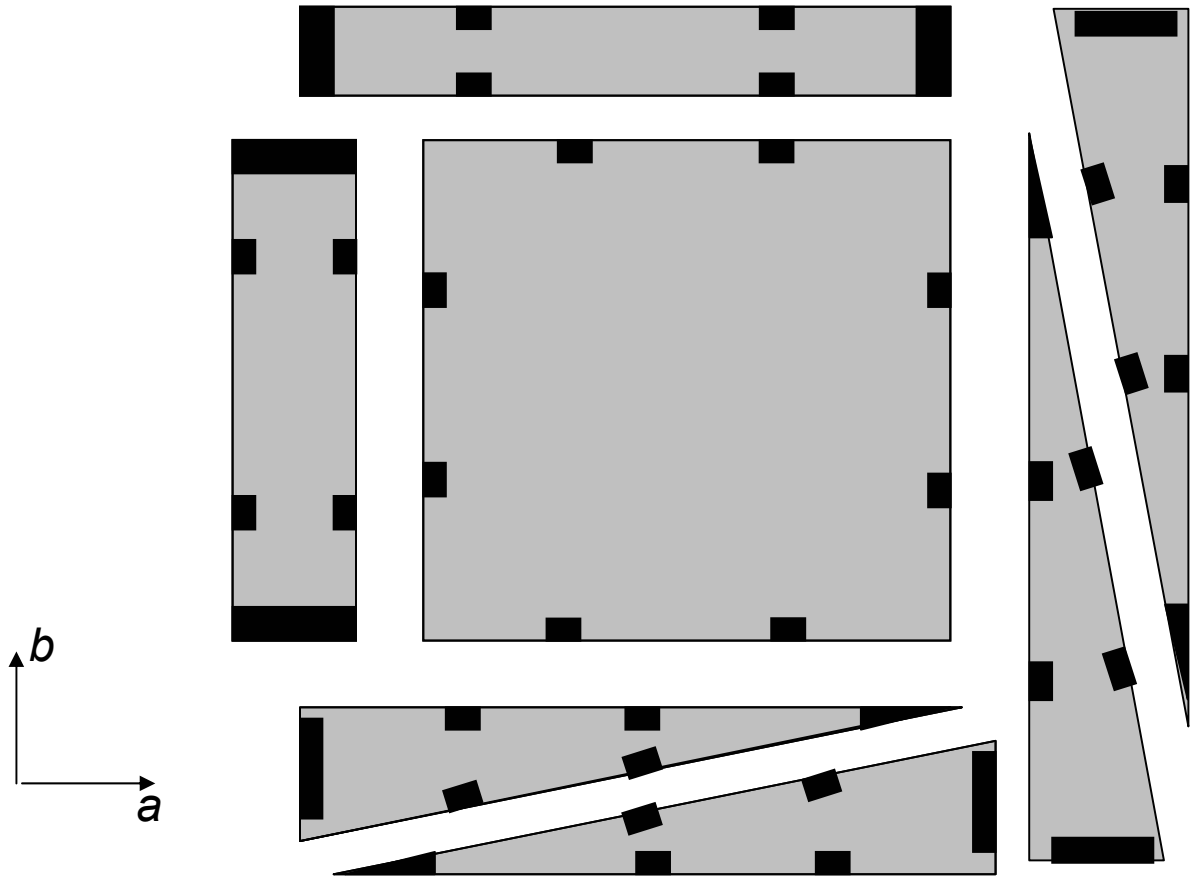

*Figure S3.1* A sketch of the crystals that were cleaved from one large square plate (initial dimension approximately 3 x 3 x 0.05 mm$^3$) for the studies of anisotropy. The rectangular plates allowed reliable measurements of $\rho_{aa}$ and $\rho_{bb}$ for all relevant combinations of directions of current versus in-plane field. The triangular plates were one check on the effect of sample shape, while with the square plate we could reproduce reproduce the $\rho_{aa}$ and $\rho_{bb}$ results by injecting and removing current from two pairs on *one crystal face and measuring voltages from pairs on the orthogonal faces. By rotating these pairs by 45° we could check the effects of passing currents at 45° to crystal axes and in-plane field component, and by rotating the samples in-plane, the effects of passing currents parallel and perpendicular to the in-plane field component but misaligned with the crystalline axes. The full dataset is far too large to reproduce here; we give this detailed description of our methods to provide confidence that we have performed a thorough study in order to draw the conclusions presented in the main text. In Fig. 1 of the main paper, the noise levels are higher than for previous measurements that we have reported on this system. This is because in order to make the multi-configuration measurements that we describe, we had to work without using the low temperature transformer-based amplification that we normally employ.*

In some cases, we observe that when we perform a doubly redundant measurement of, for example, $\rho_{aa}$ (by measuring voltage $V_{aa}$ simultaneously from both pairs of voltage contacts on the top rectangular crystal in Fig. S3.1) we see minor inconsistencies in the two results when we are inside the anomalous region. These inconsistencies disappear when the field is changed to move into the 'normal metal' regions of the phase diagram, where the voltage on the two sides coincide remarkably well. This observation could be explained in two ways. One possibility is the presence of domains of some type in the anomalous region distributed inhomogeneously in the sample. A second one is that the unusual sensitivity of the anomalous phase to sample disorder enhances any minor inhomogeneities existing in the sample that would normally go unnoticed in a Fermi liquid state.

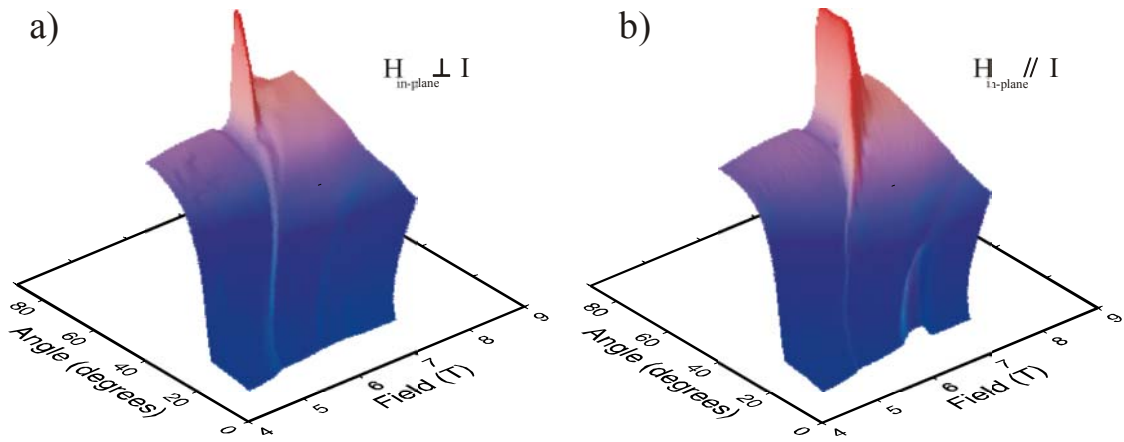

*Figure S3.2.* 3-D plots of the diagonal resistivity components $\rho_{bb}$ (panel a) and $\rho_{aa}$ (panel b) of a single crystal of $Sr_3Ru_2O_7$ as the external magnetic field is rotated from alignment along the crystalline a axis (0º) to along the crystalline c axis (90º), at a constant temperature of 100 mK. The same data with the field normalised by $H_c(\theta)$ is shown in Fig. 2.

**SOM 4**

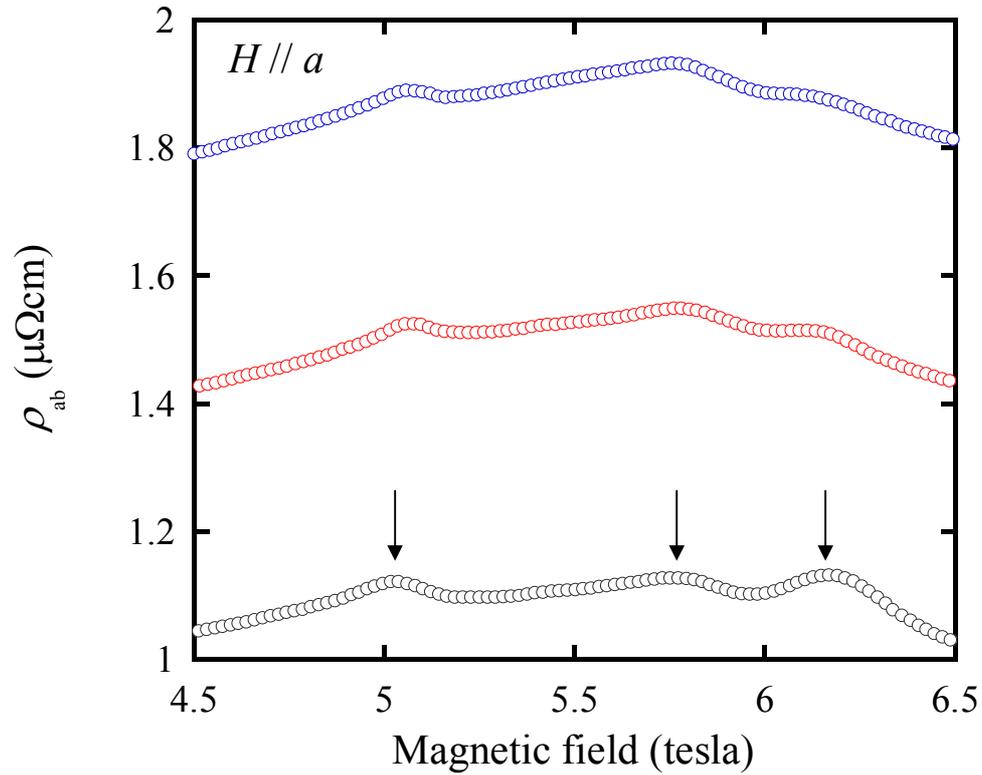

*Figure S4.1* *Magnetoresistance $\rho_{aa}$ for three $Sr_3Ru_2O_7$ single crystals with H // a at a constant temperature of 300 mK. The crystals have residual resisitivities of 0.8, 1.1 and 1.4 $\mu\Omega$cm respectively. The lowest resistivity sample shows three pronounced features, marked with arrows, at 5.0, 5.7 and 6.2 T. The two lower field features survive as the elastic scattering rate increases, but the feature at 6.2 T is rapidly washed out. Of the three features, only this one is associated with spontaneous resistive anisotropy, as can be seen in Fig. 2 of the main text.*